\documentclass[reprint,aps,prl,showpacs]{revtex4-1}
\usepackage{amsfonts}
\usepackage{psfrag}
\usepackage{graphicx}
\usepackage{cancel}
\usepackage{amsmath}
\usepackage{natbib}
\usepackage{xr}
\usepackage{hyperref}
\usepackage{dsfont} % for identity matrix
 \usepackage[bb=boondox]{mathalfa} % for bold 0

\begin{document}

\title{Quantum process tomography of linear and quadratically nonlinear optical systems}
\author{Kevin Valson Jacob\textsuperscript{1}}
\email{kjaco18@lsu.edu}
\author{Anthony E. Mirasola\textsuperscript{1,2}}
\author{Sushovit Adhikari\textsuperscript{1}}
\author{Jonathan P. Dowling\textsuperscript{1,3,4,5}}
\affiliation{\textsuperscript{1}Hearne Institute for Theoretical Physics and Department of Physics and Astronomy, Louisiana State University,
Baton Rouge, Louisiana 70803, USA }
\affiliation{\textsuperscript{2}Department of Physics and Astronomy, Rice University, Houston, Texas 77005, USA}
\affiliation{\textsuperscript{3} NYU-ECNU Institute of Physics at NYU Shanghai, Shanghai 200062, China}
\affiliation{\textsuperscript{4} CAS-Alibaba Quantum Computing Laboratory, USTC Shanghai, Shanghai 201315, China
}
\affiliation{\textsuperscript{5} National Institute of Information and Communications Technology, Tokyo 184-8795, Japan}
\begin{abstract}
A central task in quantum information processing is to characterize quantum processes. In the realm of optical quantum information processing, this amounts to characterizing the transformations of the mode creation and annihilation operators. This transformation is unitary for linear optical systems, whereas these yield the well-known Bogoliubov transformations for systems with Hamiltonians that are quadratic in the mode operators. In this paper, we propose a shot noise limited scheme for characterizing both these kinds of evolutions by employing a modified Mach-Zehnder interferometer. In order to characterize a $N$-mode device, we require $O(N^2)$ measurements. While it suffices to use coherent states for the characterization of linear optical systems, we additionally require single photons to characterize quadratically nonlinear optical systems. 
\end{abstract}
\maketitle
\section{Introduction}
Quantum process tomography is an indispensable tool in the characterization of the evolution of quantum systems. In general, the evolution of a $N$-dimensional quantum system is a completely positive trace-preserving map, which is characterized by $O(N^4)$ real parameters \cite{nielsen}. In addition to standard quantum process tomography, various schemes such as ancilla-assisted process tomography \cite{altepeter}, direct characterization of quantum dynamics \cite{lidar} and compressed sensing \cite{shabani} have been developed to characterize such maps.   

Characterizing evolutions in optical systems require a different scheme as the Hilbert space corresponding to such systems is infinite dimensional.  Several schemes have been proposed for characterizing optical systems. In Ref. \cite{lobino}, optical systems were probed with coherent states; and the results were used to predict the action of the system on an arbitrary state of light using the Glauber-Sudarshan $P$-representation. Simpler schemes are possible when we restrict our attention to linear optics.  Such systems have been found to have a variety of applications ranging from interferometry, quantum metrology \cite{giovanetti}, linear optical quantum computing \cite{dowling}, and boson sampling \cite{aaronson}. In such systems, the mode  operators evolve unitarily; and characterizing the corresponding finite dimensional unitary matrix completely specifies the evolution.

Several schemes for characterizing linear optical devices were developed in Refs. \cite{obrien,ish,melbourne}. In all these schemes, probe states are inputted into the device which is then characterized from the probabilities of specific outcomes of measurements from the output of the device. 
In Ref. \cite{obrien}, single-photon probes  were used to find the moduli of all matrix elements, and two-photon coincidence probabilities were used to find all the phases of the matrix elements of a $d$-mode unitary transformation. A similar scheme was analyzed in detail in Ref. \cite{ish}. Another approach using coherent state probes instead of single photons was developed in  Ref. \cite{melbourne}.

However, all these schemes assumed that the unitary matrix is real-bordered i.e. that the elements in the first row and first column of the matrix were real. This restricts the class of devices that we can characterize. For instance, we would not be able to characterize a single mode phase shifter by these schemes. In general, the phases in the first row and column would be relevant when either the input state is superposed across input modes or when there is further interferometry after the device. 

The restriction on the class of unitaries which could be characterized in these schemes stems from the fact that in quantum mechanics, only phase differences and not phases themselves can be measured. Thus in order to find all the phases in the transformation matrix, at least one auxillary mode must be introduced relative to which all phases can be measured. We will show that a modified Mach-Zehnder interferometer serves this purpose.

Although characterizing linear optical devices have been explored, not much attention has been given to characterizing nonlinear devices. Such devices have been shown to be useful in producing squeezed light \cite{yuen} and entangled photons \cite{shih}. Systems where the Hamiltonian is quadratic in the mode operators produce the well-known Boguliubov transformations of the mode operators \cite{lvovsky}.  We will show that the modified Mach-Zehnder interferometer can characterize such transformations also.

\section{Setup}

\begin{figure}[h]
\includegraphics[scale=0.3, angle=270]{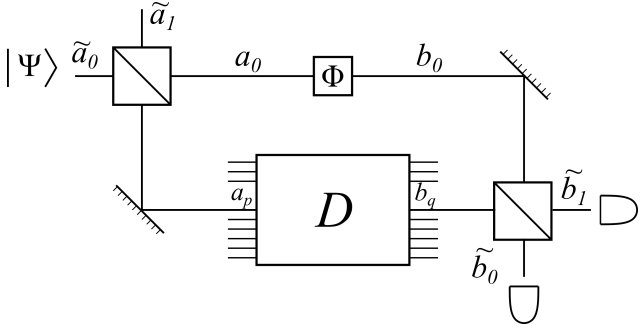}
\caption{Modified Mach-Zehnder interferometer for characterizing an unknown device $D$. The $p^{th}$ input and $q^{th}$ output modes of the device $D$ is inserted in the lower arm of the interferometer. The upper arm consists of a phase shifter which introduces a phase of 0 or $\frac{\pi}{2}$ to its input. The modes are labelled as shown. }
\end{figure}

A schematic diagram of the proposed modified Mach-Zehnder interferometer is shown as FIG. 1. It consists of two 50:50 beamsplitters, a phase shifter, the unknown device to be characterized, and photodetectors.
One of the input and output modes of the unknown device are placed in the lower arm of the interferometer; and the phase shifter is placed in the upper arm of the interferometer.
The input modes of the first beamsplitter and the output modes of the second beamsplitter are labelled as $\tilde{a}_i$ and $\tilde{b}_i$ respectively where $i = 0,1$. The first beamsplitter implements the transformation,
\begin{equation}
\left( \begin{array}{c} \hat{\tilde{a}}_0^\dagger \\ \hat{\tilde{a}}_1^\dagger \end{array} \right) = \frac{1}{\sqrt{2}}\begin{pmatrix} 1 & i \\ i & 1 \end{pmatrix}  \left( \begin{array}{c} \hat{a}_0^\dagger \\ \hat{a}_p^\dagger \end{array} \right).
\end{equation}
Here $a_p$ is the $p^{th}$ input mode of the unknown device which is coupled to the lower arm of the interferometer. The upper mode consists of a phase shifter which implements the transformation
\begin{equation}
\hat{a}_0^\dagger = e^{i\phi}\hat{b}_0^\dagger
\end{equation}
where $\phi \in \{0,\frac{\pi}{2}\}$. The output mode $b_q$ of the device and mode $b_0$ transform in the second beamsplitter as
 \begin{equation}
\left( \begin{array}{c} \hat{b}_0^\dagger \\ \hat{b}_q^\dagger \end{array} \right) = \frac{1}{\sqrt{2}}\begin{pmatrix} 1 & i \\ i & 1 \end{pmatrix}  \left( \begin{array}{c} \hat{\tilde{b}}_0^\dagger \\ \hat{\tilde{b}}_1^\dagger \end{array} \right).
\end{equation}

\section{Tomography of unitary transformations}
Consider a $N$-mode passive linear optical device where the input and output modes are labelled as $a_i$ and $b_i$ respectively where $i \in \left\lbrace1,2,...N\right\rbrace$. The input mode and output mode creation operators are related by a unitary transformation as 
\begin{equation} 
\label{unitary transformation}
\hat{a}_i^\dagger = U_{ij} \hat{b}_j^\dagger
\end{equation}
where it is implicit that the repeated index is summed over. Our aim is to fully characterize this unitary matrix.

For this, we probe it with coherent states. Consider a coherent state input in mode $\tilde{a}_0$.
The input state is 
\begin{equation}
|\Psi\rangle={\cal{D}}_{\tilde{a}_0}(\alpha)|0\rangle= e^{\alpha \hat{\tilde{a}}_0^{\dagger}-\alpha^*\hat{\tilde{a}}_0}|0\rangle
\end{equation}
where $\alpha$ is arbitrarily chosen, and ${\cal{D}}_{\tilde{a}_0}$ is the displacement operator acting on mode $\tilde{a}_0$. After the first beamsplitter, this state is 
\begin{equation}
|\Psi\rangle={\cal{D}}_{a_0}\left(\frac{\alpha}{\sqrt{2}}\right)\otimes{\cal{D}}_{a_p}\left(\frac{i\alpha}{\sqrt{2}}\right) |0\rangle
\end{equation}
After the unitary device and the phase shifter, this state is transformed as
\begin{equation}
|\Psi\rangle={\cal{D}}_{b_0}\left(\frac{e^{i\phi}\alpha}{\sqrt{2}}\right)\otimes_{j=1}^N 
\exp\left[{\frac{i\alpha U_{pj}\hat{b}_j^\dagger}{\sqrt{2}}+\frac{i\alpha U_{pj}^*\hat{b}_j}{\sqrt{2}}}\right]|0\rangle
\end{equation}
This can be rewritten as 
\begin{equation}
|\Psi\rangle={\cal{D}}_{b_0}\left(e^{i\phi}\frac{\alpha}{\sqrt{2}}\right)\otimes{\cal{D}}_{b_q}\left(\frac{i\alpha U_{pq}}{\sqrt{2}}\right)
\otimes_{j \neq q}{\cal{D}}_{b_j}\left(\frac{i\alpha U_{pj}}{\sqrt{2}}\right)|0\rangle
\end{equation}
After the final beamsplitter, the reduced state in modes $\tilde{b}_0$ and $\tilde{b}_1$ is
\begin{equation}
|\tilde{\Psi}\rangle={\cal{D}}_{\tilde{b}_0}\left(\frac{\alpha}{2}(e^{i\phi}-U_{pq})\right)\otimes{\cal{D}}_{\tilde{b}_1}\left(\frac{i\alpha}{2}(e^{i\phi}+U_{pq})\right)|0\rangle
\end{equation}
We then measure the intensity difference between the modes. This is
\begin{equation}
I_{\tilde{b}_1}-I_{\tilde{b}_0} = \left|\alpha\right|^2 \mathrm{Re}[e^{-i\phi}U_{pq}]
\end{equation}
Thus by choosing $\phi$ as 0 or $\frac{\pi}{2}$, we are able to find the real part and the imaginary part of the matrix element $U_{pq}$ respectively. By choosing $p,q\in{1,2,...N}$ we can find all the matrix elements in $O(N^2)$ measurements. This completes the characterization of the unitary matrix.

\section{Tomography of Bogoliubov transformations}
Having seen how unitary evolutions of the mode operators can be characterized, we now move on to characterizing Bogoliubov transformations. In such devices the mode operators evolve as
\begin{equation}
\hat{a}_i^\dagger=U_{ij}\hat{b}_j^\dagger + V_{ij}\hat{b}_j
\end{equation}
where $UU^\dagger-VV^\dagger = \mathds{1}$. Note that $U$ here is  unitary iff $V = 0$. Hence in general our aim is to find both $U$ and $V$. We will first find $U$, and then use that information to find $V$.

As earlier, consider a $d$-mode device  where the input and output modes are labelled as $a_i$ and $b_i$ respectively where $i \in \left\lbrace1,2,...N\right\rbrace$. For finding $U$, we use a scheme similar to the unitary case but with single photon probes. We first input a single photon in mode $\tilde{a}_0$. The state after the first beamsplitter is 
\begin{equation}
|\Psi\rangle=\left(\frac{\hat{a}_0^\dagger+i\hat{a}_p^\dagger}{\sqrt{2}}\right)|0\rangle
\end{equation}
This state is transformed to 
\begin{equation}
|\Psi\rangle=\left(\frac{e^{i\phi} \hat{b}_0^\dagger}{\sqrt{2}}+\frac{iU_{pj} \hat{b}_j^\dagger}{\sqrt{2}}\right)|0\rangle
\end{equation}
where we have noted that $\hat{b}_i|0\rangle=0 \,\forall \, i$.

The modes $b_0$ and $b_q$ transform in the beamsplitter so as to yield the final state
\begin{equation}
|\Psi\rangle=\left[\frac{(e^{i\phi}-U_{pq})}{2}\hat{\tilde{b}}_0^\dagger + \frac{(ie^{i\phi}+iU_{pq})}{2}\hat{\tilde{b}}_1^\dagger+\sum_{j\neq q}\frac{iU_{pj}{b}_j^\dagger}{\sqrt{2}}\right]|0\rangle
\end{equation}
The difference in the probabilities of measuring the photons at the output of the final beamsplitter is 
\begin{equation}
P_{\tilde{b}_1}-P_{\tilde{b}_0}=\mathrm{Re}[e^{-i\phi}U_{pq}]
\end{equation}
As earlier, by choosing $\phi$ and $p,q$, we can fully characterize the matrix $U$. 

We now need to characterize $V$. For this, we send in a coherent state probe as in the unitary case. Proceeding as earlier, we find the reduced state of modes $b_0$ and $b_q$ as
\begin{equation}
|\tilde{\Psi}\rangle={\cal{D}}_{b_0}\left(\frac{e^{i\phi}\alpha}{\sqrt{2}}\right)\otimes {\cal{D}}_{b_q}
\left(\frac{i\alpha U_{pq}}{\sqrt{2}}+\frac{i\alpha^* V_{pq}^*}{\sqrt{2}}\right)|0\rangle\
\end{equation}
For simplicity, define $\beta_{pq}=\alpha U_{pq}+\alpha^*V_{pq}^*$. 
After the final beamsplitter, this state is
\begin{equation}
|\tilde{\Psi}\rangle={\cal{D}}_{\tilde{b}_0}\left(\frac{e^{i\phi}\alpha-\beta_{pq}}{2}\right)\otimes
{\cal{D}}_{\tilde{b}_1}\left(\frac{i(e^{i\phi}\alpha+\beta_{pq})}{2}\right)
|0\rangle\
\end{equation}
In this case, the intensity difference between the outputs is
\begin{equation}
I_{\tilde{b}_1}-I_{\tilde{b}_0}=\rm{Re}[\beta_{\it{pq}}\alpha^*e^{-i\phi} ]
\end{equation}
This allows us to find $\beta_{pq}$ $\forall \, p,q$ which, can be used to find $V$ completely. Note that the above expression reduces to Eq. (10) if $V=0$. This completes the characterization of Bogoliubov transformations.

\section{Lossy devices}
Having discussed how to characterize both unitary and Bogoliubov transformations in lossless devices, we now turn our attention to lossy devices. As shown in Ref. \cite{melbourne}, if the loss is independent of the path taken by the photon in the device, then the loss can be modeled by fictitious beamsplitters. This embeds the transformation matrices of the device in a larger matrix. Akin to the model in Ref. \cite{melbourne}, we attach a fictitious beamsplitter of transmissivity $\eta_i \in [0,1]$ to the $i^{th}$ input mode of the device. This is represented in FIG. 2. 

\begin{figure}[h]
\includegraphics[scale=0.3, angle=0]{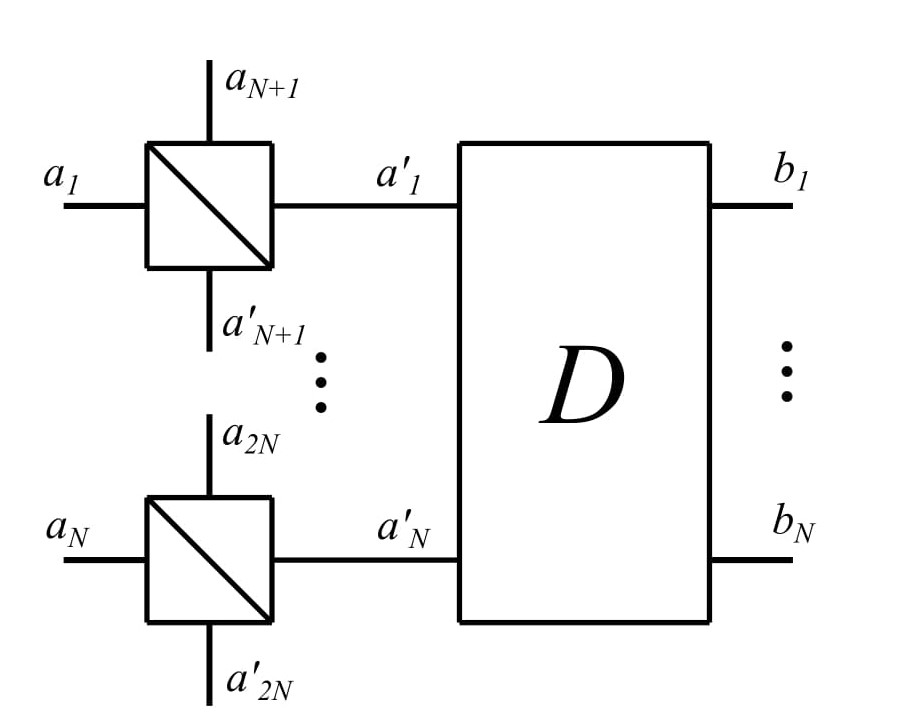}
\caption{The loss of the device is modeled by fictitious beamsplitters. The input modes of the beamsplitter are labeled $a_i$ and the input modes of the unknown device is labeled $a_i'$. The auxillary mode of the $i^{th}$ beamsplitter introduced which we have no access to is labeled $a_{N+i}$. }
\end{figure}

The fictitious beamsplitters transform the modes as

\begin{equation}
\left( \begin{array}{c}  \hat{a}_i^\dagger \\  \hat{a}_{N+i}^\dagger \end{array} \right) = \begin{pmatrix} \eta_i & -\sqrt{1-\eta_i^2} \\ \sqrt{1-\eta_i^2} & \eta_i \end{pmatrix}  \left( \begin{array}{c} \hat{a}_i^{'\dagger} \\ \hat{a}_{N+i}^{'\dagger} \end{array} \right)
\end{equation}
where $\eta_i$ is the transmissivity of the $i^{th}$ beamsplitter.
For convenience, define the diagonal matrices $\eta$ and $\tilde{\eta}$ as 
\begin{align}
\eta_{ij}&= \eta_i\delta_{ij}\nonumber \\
\tilde{\eta}_{ij}&=\sqrt{1-\eta_i^2}\delta_{ij}
\end{align}
\subsection{Unitary transformations}
We now focus on the case of a lossy unitary device. In this case, we have
\begin{equation}
\hat{a}_i^{'\dagger}=U_{ij}\hat{b}_j^\dagger
\end{equation}
Combining Eqs. (20) and (21), we obtain

\begin{equation}
\left(\begin{array}{c}  \hat{a}_1^\dagger \\ \vdots \\\hat{a}_N^\dagger \\ \hat{a}_{N+1}^\dagger \\ \vdots \\  \hat{a}_{2N}^\dagger\end{array}\right)=\begin{pmatrix}
     &   \\
     $$ \mbox{ $ (\eta U)_{N\times N}$ }$$  $$ \mbox{ $ (-\tilde{\eta}\mathds{1})_{N\times N}$ }$$  \\
     &   \\
     &   \\
     &  \\
     $$ \mbox{ $ (\tilde{\eta} U)_{N\times N}$ }$$  $$ \mbox{ $ (\eta\mathds{1})_{N\times N}$ }$$   \\
          &   \\
         
     \end{pmatrix}
    \left(\begin{array}{c}  \hat{b}_1^\dagger \\ \vdots \\\hat{b}_N^\dagger \\ \hat{a}_{N+1}^{'\dagger} \\ \vdots \\  \hat{a}_{2N}^{'\dagger}\end{array}\right)
\end{equation}
It is this $2N\times2N$ matrix that now characterizes the device. Thus, in addition to $U$, we need to find $\eta$ and $\tilde{\eta}$ also. In order to find the losses, we send in a coherent state into mode $a_i$. The state evolves as
\begin{align}
|\Psi\rangle &= {\cal{D}}_{a_i}(\alpha)|0\rangle = {\cal{D}}_{a'_i}(\eta_i \alpha)\otimes {\cal{D}}_{a'_{N+i}}(-\sqrt{1-\eta_i^2}\alpha)|0\rangle \nonumber \\ 
&=\otimes_{j=1}^N{\cal{D}}_{b_j}(\eta_iU_{ij}\alpha)\otimes {\cal{D}}_{a'_{N+i}}(-\sqrt{1-\eta_i^2}\alpha)|0\rangle
\end{align}

Then the sum of the intensities in the accessible output modes is 
\begin{equation}
I=\eta_i^2\sum_j |U_{ij}|^2|\alpha|^2=\eta_i^2|\alpha|^2
\end{equation}
From this, all $\eta_i$ and hence $\eta$ and $\tilde{\eta}$ can be found out. In order to find $U$, we proceed exactly as in the lossless unitary case. We will see that Eq. (10) will be modified to read
\begin{equation}
I_{\tilde{b}_1}-I_{\tilde{b}_0} = \left|\alpha\right|^2 \mathrm{Re}[e^{-i\phi}\eta_p U_{pq}]
\end{equation}
from which we can now find $U$. Thus the lossy unitary device can be characterized.

\subsection{Bogoliubov transformations}

We now move on to lossy devices that implement Bogoliubov transformations. In such devices, the mode operators evolve as 
\begin{equation}
\hat{a}_i^{'\dagger}=U_{ij}\hat{b}_j^\dagger + V_{ij}\hat{b}_j
\end{equation}
Modeling the loss as earlier, the full transformation becomes

\begin{align}           %start alignment between state and observation equation 
%State Equation
%\begin{split}           %split the observation equation in two so it fits on the page
\left(\begin{array}{c}  \hat{a}_1^\dagger \\ \vdots \\\hat{a}_N^\dagger \\ \hat{a}_{N+1}^\dagger \\ \vdots \\  \hat{a}_{2N}^\dagger\end{array}\right)&=\begin{pmatrix}
    & \\
    $$ \mbox{ $ (\eta U)_{N\times N}$ }$$ $$ \mbox{ $ (-\tilde{\eta}\mathds{1})_{N\times N}$ }$$   \\
    & \\
     & \\
    &  \\
   $$ \mbox{ $ (\tilde{\eta} U)_{N\times N}$ }$$ $$ \mbox{ $ (\eta\mathds{1})_{N\times N}$ }$$   \\
       &  \\
     \end{pmatrix}
    \left(\begin{array}{c}  \hat{b}_1^\dagger \\ \vdots \\\hat{b}_N^\dagger \\ \hat{a}_{N+1}^{'\dagger} \\ \vdots \\  \hat{a}_{2N}^{'\dagger}\end{array}\right) \notag \\ %split here
&\quad +  
\begin{pmatrix}
    &  \\
    $$  \mbox{ $ (\eta V)_{N\times N}$ }$$ & &$$ \mbox{ $ -\mathbb{0}_{N\times N}$ }$$   \\
    & \\
    & \\
    &  \\
    $$ \mbox{ $ (\tilde{\eta} V)_{N\times N}$ }$$ & &$$ \mbox{ $ \mathbb{0}_{N\times N}$ }$$  \\
    & \\
     \end{pmatrix}
    \left(\begin{array}{c} \hat{b}_1 \\ \vdots \\\hat{b}_N \\ \hat{a}'_{N+1} \\ \vdots \\  \hat{a}'_{2N}\end{array}\right)
\end{align}

Thus we have to find $U$, $V$, $\eta$, and $\tilde{\eta}$ in order to fully characterize this device. In order to find $\eta$ we send in a single photon in mode $a_i$. The probability that the photon will be detected in any of the accessible output modes in $|\eta_i|^2$. Thus $\eta$ and $\tilde{\eta}$ can be found out. To find $U$, we proceed exactly as in the case of lossless Bogoliubov transformations so that Eq. (15) will be modified to
 \begin{equation}
P_{\tilde{b}_1}-P_{\tilde{b}_0}=\mathrm{Re}[e^{-i\phi}\eta_pU_{pq}] 
\end{equation}
from which $U$ can be found out. 
Similarly Eq. (18) will be modified to 
\begin{equation}
P_{\tilde{b}_1}-P_{\tilde{b}_0}=\mathrm{Re}[e^{-i\phi}\alpha^*\eta_p\beta_{pq}]
\end{equation}
which enable us to find $V$. This completes the characterization of lossy Bogoliubov transformations

\section{Conclusion}
We have shown that a modified Mach-Zehnder interferometer can characterize both unitary and Bogoliubov transformations. As we have used coherent states and single photons in our scheme, the sensitivity of our scheme is limited by shot noise.

\begin{acknowledgments}
The authors would like to acknowledge support from the the Air Force Office for Scientific Research, the Army Research Office, the Defense Advanced Projects Agency, the National Science Foundation, and the Northrop Grumman Corporation. The authors thank Mark M. Wilde, Xiaoting Wang, and Chenglong You for helpful discussions and comments on this work. 
\end{acknowledgments}

\bibliographystyle{apsrev}

\begin{thebibliography}{100}

\bibitem{nielsen} M. A. Nielsen and I. L. Chuang, \textit{Quantum Computation and Quantum Information} (Cambridge university Press, Cambridge, UK, 2000).

\bibitem{altepeter} J. B. Altepeter, D. Branning, E. Jeffrey, T. C. Wei, P. G. Kwiat, R. T. Thew, J. L. O’Brien, M. A. Nielsen, and A. G. White,
Phys. Rev. Lett. {\bf{90}}, 193601 (2003).

\bibitem{lidar} M. Mohseni and D. A. Lidar, Phys. Rev. Lett. {\bf{97}}, 170501 (2006).

\bibitem{shabani}A. Shabani, R. L. Kosut, M. Mohseni, H. Rabitz, M. A. Broome, M. P. Almeida, A. Fedrizzi, and A. G. White, Phys. Rev. Lett. {\bf{106}}, 100401 (2011).

\bibitem{lobino} M. Lobino, D. Korystov, C. Kupchak, E. Figueroa, B. C. Sanders, and A. I. Lvovsky, Science, {\bf{322}}, 563 (2008).

\bibitem{giovanetti} V. Giovannetti, S. Lloyd, and L. Maccone, Phys. Rev. Lett. {\bf{96}}, 010401 (2006).

\bibitem{dowling}P. Kok, W. J. Munro, K. Nemoto, T. C. Ralph, J. P. Dowling, and G. J. Milburn, Rev. Mod. Phys. {\bf{79}}, 135 (2007).

\bibitem{aaronson} A. Arkhipov and S. Aaronson,  Proc.
ACM STOC (New York), (2011).

\bibitem{obrien} A. Laing and J. O'Brien, arXiv:1208.2868 (2012).

\bibitem{ish} I. Dhand, A. Khalid, H. Lu, and B. C. Sanders,
J. Opt. {\bf{18}}, 035204 (2016).


\bibitem{melbourne} S. Rahimi-Keshari, M. A. Broome, R. Fickler,
A. Fedrizzi, T. C. Ralph, and A. G. White, Optics express {\bf{21}}, 13450 (2013).

\bibitem{yuen} H. P. Yuen, Phys. Rev. A {\bf{13}}, 2226 (1976).

\bibitem{shih} Y. Shih, Rep. Prog. Phys. {\bf{66}}, 1009 (2003).

\bibitem{lvovsky} A. I. Lvovsky, arXiv:1401.4118 (2014).
\end{thebibliography}

\end{document}